\documentstyle[preprint,tighten,floats,aps]{revtex}
\begin{document}
 
 
\title{\large\bf Chiral Symmetry, Quark Mass, and Scaling of the \\Overlap Fermions}
 
\author{S.J. Dong$^{a}$, F.X. Lee$^{b,c}$,
K.F. Liu$^{a}$, and J.B. Zhang$^{a,d}$}
 
\address
{$^{a}$Dept. of Physics and Astronomy, University of Kentucky,
Lexington, KY 40506\\
$^{b}$ Center for Nuclear Studies, Dept. of Physics, George Washington University,
\\Washington, DC 20052 \\
$^{c}$Jefferson Lab, 12000 Jefferson Avenue, Newport News, VA 23606 \\
$^{d}$ Zhejiang Institute of Modern Physics, Zhejiang University, Hangzhou 310027,
China}
 
 
\maketitle
\vskip 12pt
 
\begin{abstract}
The chiral symmetry relation and scaling of the overlap fermions are
studied numerically on the quenched lattices at 3 couplings with about
the same physical volume. We find that the generalized Gell-Mann-Oakes-Renner
relation is satisfied to better than 1\% down to the smallest quark mass
at $m_0a = 0.006$. We also obtain the quark mass from the PCAC relation and
the pseudoscalar masses. The renormalization group
invariant quark mass is shown to be fairly independent of scale.
The $\pi$ and $\rho$ masses at a fixed
$m_{\pi}/m_{\rho}$ ratio indicate small $O(a^2)$ corrections. It is
found that the critical slowing down sets in abruptly at a very small quark
mass close to those of the physical $u$ and $d$ quarks.
 
\end{abstract}
 
Recent advances in chiral fermion formulation on the lattice hold a great promise 
in implementing chiral fermions for QCD at finite lattice spacing~\cite{neu00}. 
There are distinct advantages over the previous formulations. For example, the Wilson 
fermion breaks chiral symmetry at finite lattice spacing $a$ and, therefore, the task 
of extrapolating the Monte Carlo results to the continuum and
chiral limits requires fine tuning and is often difficult. In this case, one could not
expect low energy theorems to be reproduced at finite lattice spacing. Similarly,
the staggered fermion can only be formulated with four flavors with an $U(1)$ subgroup
of the flavor non-singlet chiral symmetry. At finite lattice
spacing where the numerical calculations are carried out, the flavor symmetry
is broken. Furthermore, the anomalous chiral Ward identity does not hold
for these fermions unless at the continuum limit. This makes the
analysis of anomaly on the lattice rather unclear. There is no unambiguous
identification of the fermion zero modes with the topology of the background
gauge field. These difficulties have rendered the studies of
low energy phenomenology of QCD on the lattice unsettling.
 
 The picture has been altered dramatically with the advent of Neuberger's
overlap fermion~\cite{neu98} which is derived from the the overlap 
formalism~\cite{nn95}. All of the above-mentioned impediments can be 
avoided pending pristine numerical simulation.
It is shown to have correct anomaly and exact chiral symmetry on the lattice
~\cite{nn95,nn93} and there are no order $a$ artifacts~\cite{knn97}.  
The overlap fermion has a compact form in four dimension and is
easily employed to derive low energy theorems, chiral symmetry relations, and
anomaly at finite lattice spacing. In this letter, we shall study the overlap
fermion numerically. We test
chiral symmetry via the Gell-Mann-Oakes-Renner relation at finite lattice spacing
and check the scaling behavior of the $\pi$ and $\rho$ masses. The
preliminary results were reported earlier~\cite{ldl00}.
We also obtain the quark mass from the Chiral Ward identity and verify
that it is free from additive renormalization and the renormalization group invariant
quark mass is indeed independent of scale.
 
Neuberger's Dirac operator has the following form for the massive
case~\cite{neu98a}
\begin{equation}  \label{neu}
D(m_0)=  1 + \frac{m_0a}{2} + (1 - \frac{m_0a}{2} ) \gamma_5 \epsilon (H),
\end{equation}
where $\epsilon (H) = H /\sqrt{H^2}$ is the matrix sign function of H which we
take to be the Hermitian Wilson-Dirac operator, i.e.
$H = \gamma_5 D_w$. Here $D_w$ is the usual Wilson fermion operator, except
with a negative mass parameter which corresponds to $\kappa_c < \kappa < 0.25$.
We take $\kappa = 0.19$ in our calculation.
The massless operator $D(0)$ is shown~\cite{neu98b} to satisfy the Ginsparg-Wilson 
relation~\cite{gw82}:
$\{\gamma_5, D(0)\} = D(0)\gamma_5 D(0)$. The bare mass parameter $m_0$ is
proportional to the quark mass without additive constant which we shall verify.
 
There are several numerical approaches to approximate the sign
function $\epsilon (z)$~\cite{neu98c,chi98,ehn99a,hjl99}.
We adopt the optimal rational approximation~\cite{ehn99a} with a ratio of
polynomials of degree 12 in the Remez algorithm. We find the error to the approximation
of $\epsilon (z)$
to be within $10^{-5}$ in the range [0.02, 2] of the argument $z$.
In the range [0.0005, 0.02], the error can be as large as 1\%. To improve the accuracy of
$\epsilon (H)$
and hence the chiral symmetry property, the smallest 10 to 20 eigenvalues of $H^2$
with eigenvalues of $|H|$ less than 0.04 are projected out for exact evaluation
of the sign function from these eigenstates~\cite{ehn99a}. This has the added
benefit of reducing the number of iterations for the multi-mass conjugate gradient
inversion of $H^2 + c_i$ in the inner loop by a factor of 3.5 or so.
We checked the unitarity of the matrix $V = \gamma_5 \epsilon (H)$. For $Vx = b$, we
find $|x^{\dagger}x - b^{\dagger}b| \sim 10^{-9}$. Since $V$ is unitary we
exploit the identity $(1 + V^{\dagger})(1 + V) = 2 + V + V^{\dagger}$ in order to use the
conjugate gradient algorithm on the hermitian matrix $V + V^{\dagger}$ instead of
$V^{\dagger} V$ which has a higher condition number. Furthermore, since
$[V + V^{\dagger}, \gamma_5] = 0$, one can use a chiral source, {\it i.e.}
$\gamma_5 b = \pm b$
to save one matrix multiplication~\cite{ehn99b} per iteration. In the present study, we
consider three lattices: 
$6^3 \times 12$ at $\beta = 5.7$, $8^3 \times 16$ at $\beta =5.85$, and
$10^3 \times 20$ at $\beta = 6.0$, which have about the same physical volume.
With residuals at $10^{-7}$, the inner loop takes typically $\sim 200 - 250$ iterations and
is almost independent of the lattice volume. The outer loop takes $\sim 40 - 110$
iterations from the small to large lattice volumes. Even for topological
sectors with charge $Q \neq 0$, we find the critical slowing down is much milder
than that of the Wilson fermion and we found no exceptional configurations.
The critical slowing down sets in quite abruptly when $m_0 a < 0.006$ for the
sector with topology. This is already very close to the physical $u$ and $d$ masses.

The lattice chiral symmetry is reflected in the generalized
Gell-Mann-Oakes-Renner (GOR) relation
\begin{equation}   \label{gor}
m_0a\int d^4x \langle \pi(x) \pi(0)\rangle = 2 \langle\bar{\psi}\psi\rangle.
\end{equation}
Since it is satisfied for each quark mass, lattice volume and spacing, configuration
by configuration, and for each source~\cite{nn95,ehn99b,cha99}, it serves as an 
economic test of chiral symmetry. Here $\pi(x) = \bar{\psi}\gamma_5 (\tau/2)\psi$ is 
the pion interpolation field and the quark propagator is the external 
one~\cite{neu98a,chi99} with 
$D_c^{-1} = (1 - m_0a/2)^{-1} [D^{-1}(m_0)- 1/2]$. Alternatively, 
one can use the bilinears $(1 - m_0a/2)^{-1} \bar{\psi}\Gamma(1 - D/2)\psi$ for the 
operators and $D^{-1}$ as the propagator.
We utilize this relation in Eq.~(\ref{gor}) as a check of the numerical implementation
of the Neuberger operator. We find that for the lattices we consider
the GOR relation is satisfied very
well (to within $10^{-3}$) all the way down to the smallest mass $m_0 a = 2 \times
10^{-4}$ for the $Q = 0$ sector.
For the $Q \neq 0$ sector, the presence of zero modes demands higher
precision for the approximation of $\epsilon (H)$. 
\begin{figure}[ht]
\vspace{5.1cm}
\includegraphics{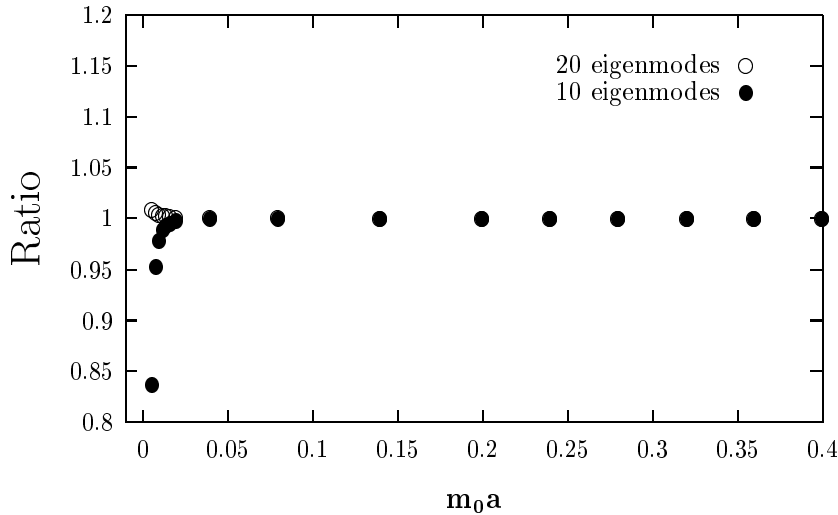}
\vspace{1.5cm}
\caption{Ratio of the right to left side of Eq. (\ref{gor}) 
for a configuration with topology. The symbols $\bullet/\circ$ indicate the case
with projection of 10/20 smallest eigenmodes.}
\label{z2qn0_r}
\end{figure}
For example, we show in Fig.~\ref{z2qn0_r} the ratio of the right to left side of
Eq.~(\ref{gor}) for a typical configuration with topology on the
$6^3 \times 12$ lattice at $\beta = 5.7$ as a function of the quark mass $m_0a$.
When only 10 smallest eigenmodes of $H^2$ are
projected, we see that the ratio deviates from one for small quark masses.
For the smallest mass $m_0a = 0.006$, it may deviate as much as 16\%. The situation
is considerably improved when 20 smallest eigenmodes are included where the deviation
is reduced to 1\%. Our result with the overlap is appreciably better than the
domain-wall fermion case when the size of the fifth dimension is limited to
$L_s = 10$ to 48~\cite{che98}. In the latter case, the ratio deviates from unity by
$\sim 55\%$ for $L_s = 10$ and $\sim 15\%$ for $L_s = 48$ at the quark mass $m_fa = 0.02$
which is about 3 times larger than the mass of $m_0a = 0.006$ in our study
where the deviation is at the 1\% level.

\begin{figure}[tbh]
\includegraphics{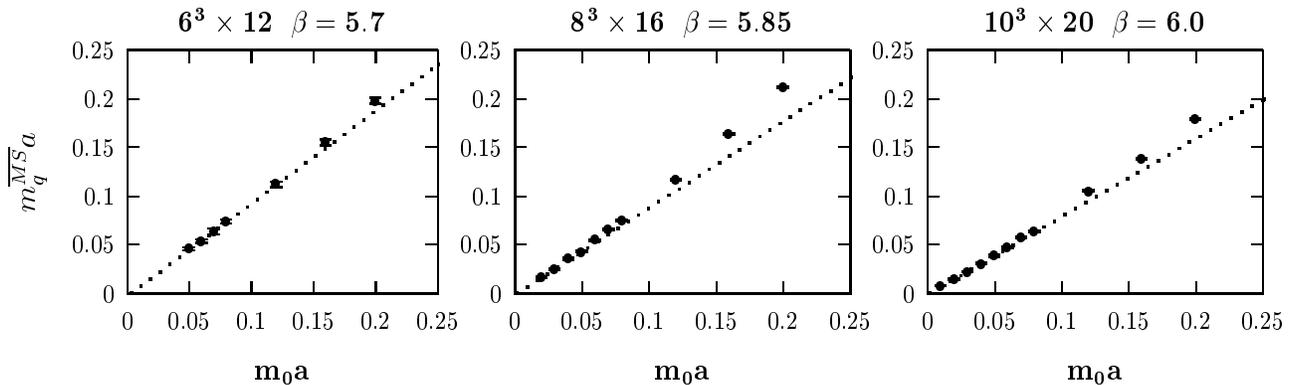}
\vspace{5.0cm}
\caption{Renormalized quark mass vs. the bare quark mass on the three lattices.}
\label{rqmass}
\end{figure}

The average of $u$ and $d$ quark masses in the $\overline{MS}$ scheme at the
renormalization scale $\mu$ is
obtained from the axial Ward identity $Z_A\partial_{\mu} A_{\mu} = 2 Z_S^{-1}m_0
Z_P P$ via the ratio of the matrix elements
\begin{equation} \label{qmass}
m_q^{\overline{MS}} (\mu) = Z_S^{-1}(\mu) m_0 = \lim_{t \rightarrow \infty}
\frac{Z_A(\mu)}{Z_P(\mu)} \frac{\sum_{\vec{x}}\langle 0|\nabla_t
A_4(x)|\pi(0)\rangle}{2\sum_{\vec{x}}\langle 0|P(x)|\pi(0)\rangle}
\end{equation}
where $A_{\mu}= \bar{\psi}i\gamma_{\mu}\gamma_5(\tau/2)\psi$ and
$P = \bar{\psi}i\gamma_5(\tau/2)\psi$. The resumed cactus diagram of
one-loop calculation of $Z_A$ and $Z_P$~\cite{afp00} are used in
Eq. (\ref{qmass}). It turns out that the ratio $Z_A/Z_P$ is very close
to unity for $\mu  = 1/a$. The renormalized quark mass $m_q^{\overline{MS}}a$
defined in Eq. (\ref{qmass}) is plotted in Fig.~\ref{rqmass}
against the bare mass $m_0a$ for the three lattices.
 
We first observe that the renormalized quark mass does not have an additive part
due to the lattice chiral symmetry. The linear fit including the smallest 5, 7, and
8 quark masses for $\beta = 5.7, 5.85$, and 6.0 respectively shows that the intercepts
are of the order of $10^{-3}$ and are consistent with zero. This is a distinct advantage
over the Wilson fermion whose quark mass is subject to additive renormalization which
depends on the gauge configuration. From the slope we can determine the non-perturbative
$Z_S^{-1}$ which are 0.95(5), 0.89(5) and 0.80(1) for $\beta = 5.7, 5.85$, and 6.0.
These are within 14\% from the perturbative calculation which are 0.818, 0.824, and
0.829 respectively~\cite{pan00}. The renormalization group invariant quark mass is defined
as the integration constant of the evolution equation such that
\begin{equation}
m_q^{RGI} = \Delta Z_S (\mu) m_q^{\overline{MS}}(\mu) = \Delta Z_S (\mu) Z_S^{-1}(\mu) m_0.
\end{equation}
Using the four-loop calculation of $\Delta Z_S (\mu)$ in the $\overline{MS}$
scheme~\cite{gho99}, we obtain the product of $\Delta Z_S (\mu) Z_S^{-1}(\mu)$
which should be scale invariant. The results from the 3 lattices give
1.12(1), 1.17(6), and 1.14(5) for  $\beta = 5.7, 5.85$, and 6.0. They are indeed independent
of scale within errors. From the pion mass fit to be discussed later,
we determine $m_0 = 6.8(5)$ MeV from the physical pion mass on the $\beta = 6.0$ lattice.
This gives $m_q^{RGI} = 7.6(6)$ MeV for the average of $u$ and $d$ quark masses, which
in turn gives $m_q^{\overline{MS}}(\mu = 2 {\rm GeV}) = 5.5(5)$ MeV. The lattice scale is
set by the $r_0$ as derived from the static quark potential~\cite{gsw98}.
We should point out that this quark mass should not be taken literally, since the volume
is quite small. The primary purpose of the
present study is to verify that the quark mass extracted from the overlap fermion action
does not have additive renormalization and that $m_q^{RGI}$ is indeed scale invariant.
 
\begin{figure}[h]
\vspace*{10.0cm}
\includegraphics{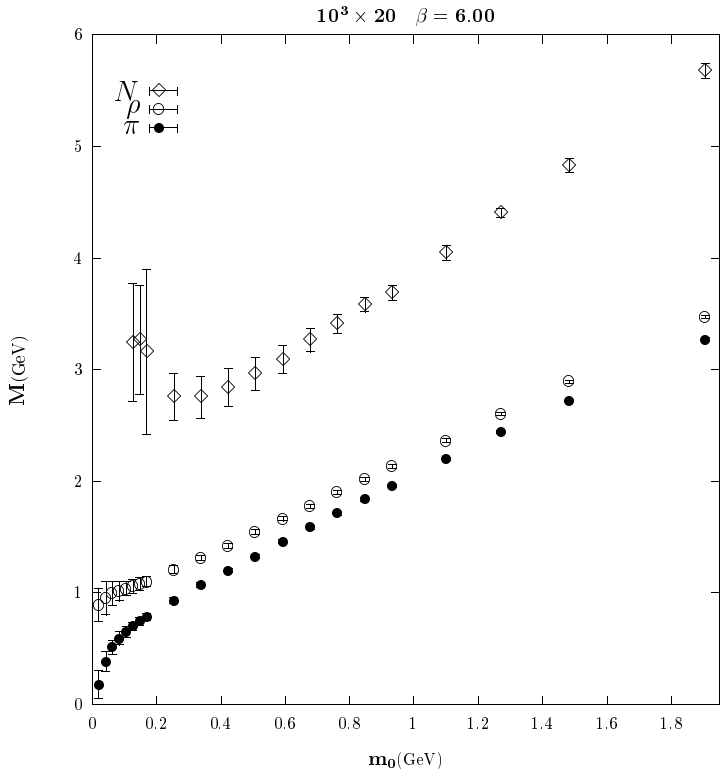}
\vspace{0.3cm}
\caption{Masses of $\pi, \rho$ and $N$ on the  $10^3 \times 20$ lattice at
$\beta = 6.0$ vs. $m_0$. The scale $r_0$ is used for conversion to physical units.}
\label{mass_10x20}
\end{figure}

 
Next we look at hadron masses. The results for the $\pi, \rho$ and nucleon masses on
the $10^3 \times 20$ lattice at $\beta = 6.0$ are given in Fig.~\ref{mass_10x20}. They
are obtained by a single exponential fit with covariance matrix.
 
\begin{figure}[tbh]
\vspace*{10.0cm}
\includegraphics{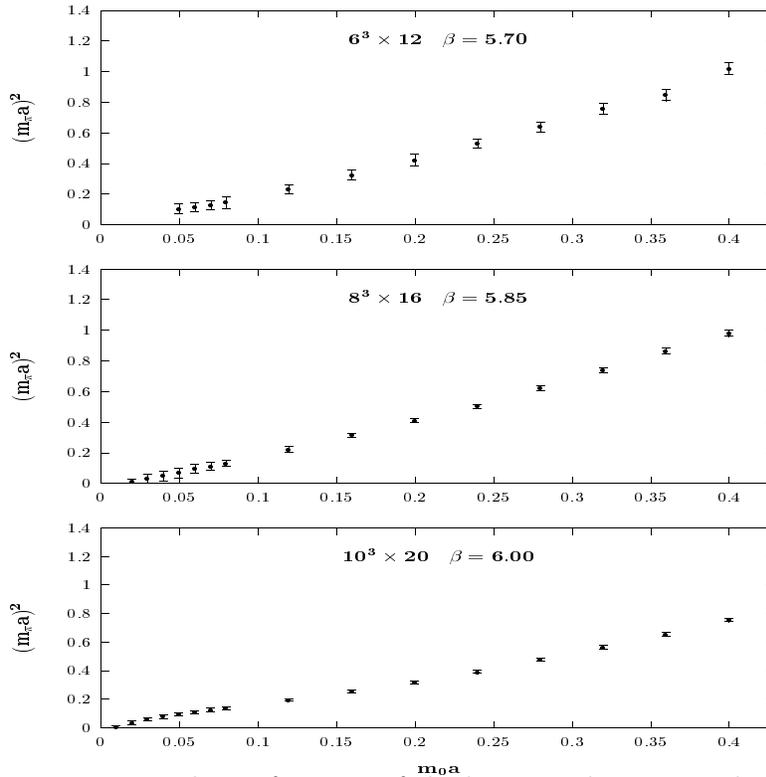}
\caption{Pion mass squared as a function of the bare quark mass on the three lattices.}
\label{pimass}
\end{figure}

We see clearly that the nucleon mass
suffers from the finite volume effect when $m_0 $ is smaller than 0.4 GeV.
Although the $\rho$ and $\pi$ masses appear not affected as much, there is the worry
that the finite volume effect is present when one considers the pion mass behavior as a
function of the quark mass $m_0$. In the finite size scaling region where the lattice
size is much smaller than the pion Compton wavelength ($L << 1/m_{\pi}$), the eigenvalue
distribution and the chiral condensate can be described by the universal function derived
in the chiral random matrix theory~\cite{sv93}. While the chiral condensate of the overlap
fermion in this
region has been studied~\cite{deh00}, we shall concentrate on the region $L > 1/m_{\pi}$
where the chiral perturbation analysis is expected to apply.
Plotted in Fig.~\ref{pimass} are the pseudoscalar meson mass squared ($m_P^2a^2$) as a
function of $m_0a$ for the three lattices for those pseudoscalar mesons whose
Compton wavelengths are less than the respective lattice size.
 
We fit them with a form suggested by the quenched chiral perturbation theory~\cite{sbg92}
\begin{equation}  \label{chi_log}
m_P^2 a^2 = 2A m_0 a^2\{1 -\delta \ln(2Am_0/\Lambda_{\chi}^2)\} + 4B m_0^2.
\end{equation}
We find that for $\Lambda_{\chi}$ in between 0.6 GeV and 1.4 GeV, the chiral
log $\delta$ for the $\beta = 6.0$ case is in the range of 0.20(2) to 042(14) which
is slightly larger than what is obtained from the Wilson fermion~\cite{abb99}.
The fit is quite stable with small $\chi^2$ ($\chi^2/DF < 0.1$) and is fairly insensitive
to the range of the fitted quark masses shown in Fig.~\ref{pimass}. On the other hand, the
fit for $\beta = 5.85$ and 5.7 lattices shows that $\delta$ is consistent with zero (the
typical value ranges from -0.04(21) to -0.34(48)) and the result is more sensitive to the
range of quark masses that are fitted. We are not able to draw any definite conclusion from
these inconsistent results, except to point out that perhaps this is due to the finite
volume effect. After all, the physical volumes of the lattices we use
are quite small ($L \sim 1 fm$). Another possibility is that the smaller
quark mass cases may already fall into the finite size scaling region where the pion mass
may have a different behavior than prescribed in Eq. (\ref{chi_log}). This will be
studied elsewhere.
 
\begin{figure}[t]
\includegraphics{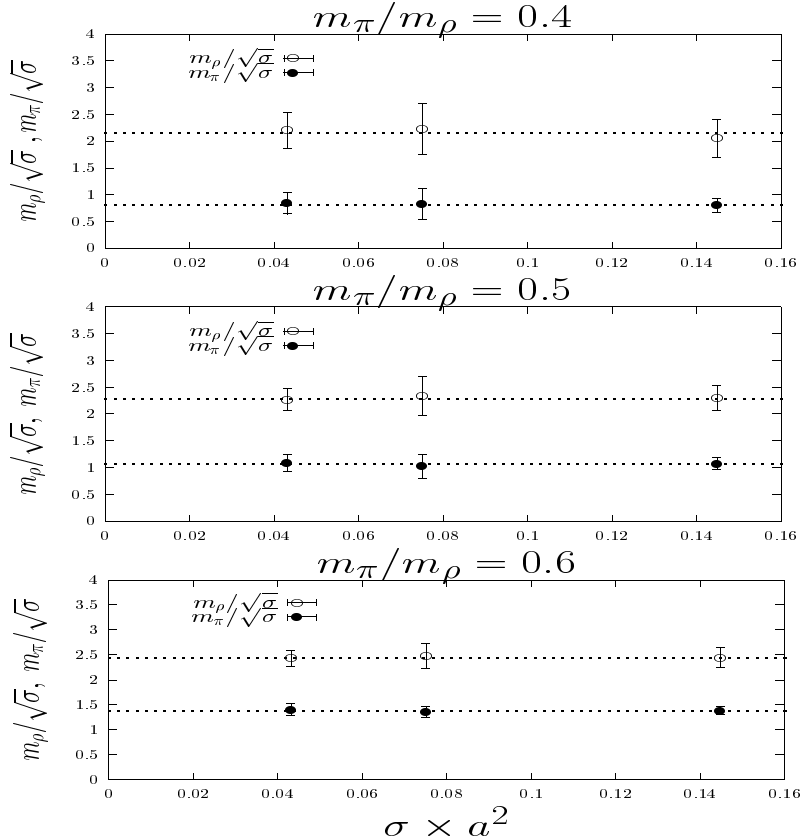}
\vspace{10.5cm}
\caption{Pion and rho masses in units of $\sqrt{\sigma}$ 
on the three lattices are plotted against $\sigma a^2$.
They are given at three different $m_{\pi}/m_{\rho}$ ratios.
The dashed line is the fit with a constant.}
\label{scale}
\end{figure}

Finally, we examine the scaling of $\rho$ and $\pi$ masses. Since $m_{\pi}^2 a^2$
and $m_{\rho}a$ are fairly linear in $m_0a$, we fit them with
$m_{\rho}a = A + B m_0a$ and $m_{\pi}^2 a^2 = C m_0a + D m_0^2a^2$ for simplicity.
The fits are decent with $\chi^2/DF < 1$ for the three lattices and the full
range of quark masses in Fig.~\ref{pimass}.
From the fits of $m_{\rho}a$ and $m_{\pi}^2 a^2$, we determine $m_0$ for which
the ratio of $m_{\pi}/m_{\rho} = 0.4, 0.5, 0.6$ and plot in Fig.~\ref{scale} the
corresponding $m_{\rho}$ and $m_{\pi}$ in units of $\sqrt{\sigma}$ as a function of
$\sigma a^2$, where $\sigma$ is the string tension. The errors on the vector
and pseudoscalar masses are determined from interpolating the data from the
neighboring quark masses. It is known that the overlap operator does not have $O(a)$
artifacts~\cite{knn97}. It appears from Fig.~\ref{scale} that even the $O(a^2)$
errors are small.
 
To conclude, we find that when the matrix sign function $\epsilon (H)$ in the overlap
fermion is well approximated, the promised chiral symmetry at finite lattice
spacing and the scaling of the renormalization group invariant quark mass and hadron masses
are manifested in the present numerical calculation. One drawback of the overlap fermion is
the large numerical overhead in the present algorithm. But, the stake of
being able to implement chiral symmetry at finite lattice spacing is high.
Furthermore, the unexpected feature of being able to push the critical slowing down to
close to the physical $u$ and $d$ quark masses has the advantage of being able to
study the correct chiral behavior. The nice
scaling result is encouraging for controlling the continuum extrapolation and
may afford the possibility of working at relatively large lattice spacings.
For the moment, the study is limited to small volumes. As long as one can extend it
to large volumes, it appears that one will be at a stage of putting all the systematic
errors of at least the quenched approximation under control.

This work is partially supported by
DOE Grants DE-FG05-84ER40154 and DE-FG02-95ER40907.
We thank R. Edwards for sharing his experience in implementing the
sign function solver. We also thank H. Neuberger and T.W. Chiu for many
stimulating discussions, R. Narayanan for communication, and H. Panagopoulos for 
calculating the renormalization constants for us.

\end{document}